\documentclass[a4paper,11pt]{article}
\usepackage{a4wide,amsmath,amssymb,bbm}
\usepackage[english]{babel}
\usepackage[utf8]{inputenc}

\usepackage{color}

\usepackage{hyperref,enumerate}
\hypersetup{
    colorlinks,
    linkcolor={black},
    citecolor={black},
    urlcolor={black}
}

\DeclareMathOperator{\tr}{tr}
\renewcommand{\o}{\overline}
\renewcommand{\u}{\underline}

\makeatletter

\@addtoreset{equation}{section} \makeatother

\begin{document}
\setcounter{page}{0}

\hfill
\vspace{30pt}

\begin{center}
{\huge{\bf\boldmath $O(D,D)$ and the string $\alpha'$ expansion:\\ An obstruction}}

\vspace{80pt}

Stanislav Hronek  \ \ and \ \ Linus Wulff

\vspace{15pt}

\small {\it Department of Theoretical Physics and Astrophysics, Faculty of Science, Masaryk University\\ 611 37 Brno, Czech Republic}
\\
\vspace{12pt}
\texttt{436691@mail.muni.cz, wulff@physics.muni.cz}\\

\vspace{80pt}

{\bf Abstract}
\end{center}
\noindent
Double Field Theory (DFT) is an attempt to make the $O(d,d)$ T-duality symmetry of string theory manifest, already before reducing on a $d$-torus. It is known that supergravity can be formulated in an $O(D,D)$ covariant way, and remarkably this remains true to the first order in $\alpha'$. We set up a systematic way to analyze $O(D,D)$ invariants, working order by order in fields, which we carry out up to order $\alpha'^3$. At order $\alpha'$ we recover the known Riemann squared invariant, while at order $\alpha'^2$ we find no independent invariant. This is compatible with the $\alpha'$ expansion in string theory. However, at order $\alpha'^3$ we show that there is again no $O(D,D)$ invariant, in contradiction to the fact that all string theories have quartic Riemann terms with coefficient proportional to $\zeta(3)$. We conclude that DFT and similar frameworks cannot capture the full $\alpha'$ expansion in string theory.

\clearpage
\tableofcontents

\section{Introduction}
When we compactify string theory on a $d$-torus, $T^d$, T-duality leads to an $O(d,d)$ symmetry group (for superstrings the group is larger but contains $O(d,d)$). While the exact symmetry is $O(d,d;\mathbbm Z)$, the massless sector of the theory displays a global $O(d,d;\mathbbm R)$ symmetry \cite{Meissner:1991zj,Sen:1991zi}. Double Field Theory (DFT) \cite{Siegel:1993th,Hull:2009mi,Hohm:2010jy} is an attempt to make this duality symmetry manifest already \emph{before} compactification.\footnote{See also \cite{Duff:1989tf,Tseytlin:1990nb,Tseytlin:1990va} for earlier work. Nice reviews include \cite{Aldazabal:2013sca,Berman:2013eva,Hohm:2013bwa}.} To achieve this one doubles the dimension, replacing $x^m\rightarrow X^M=(\tilde x_m,\,x^m)$, where the doubled coordinates $X^M$ are rotated by $O(D,D)$ ($D=10$ for superstrings and $D=26$ for the bosonic string). The equivalence to the original $D$-dimensional description is ensured by imposing an $O(D,D)$ invariant section condition which effectively eliminates half of the coordinates. Here we will work in a setting where the section condition is solved in the standard way. Namely we take $X^M=(0,\,x^m)$ and correspondingly $\partial_M=(0,\,\partial_m)$. This is equivalent to working in generalized geometry where the tangent and cotangent bundle of the manifold are unified \cite{Hitchin:2004ut,Gualtieri:2007ng}. In particular the structure group will consist of two copies of the Lorentz group $SO(D-1,1)$.

DFT leads to a reformulation of supergravity, which is very useful for certain types of questions. However, it is not so clear whether this formalism can work when $\alpha'$-corrections are included (unless one restricts to the compactified situation). One reason to expect that it would be difficult to describe higher derivative corrections is the absence of an $O(D,D)$ covariant Riemann tensor \cite{Hohm:2011si}. Conversely, if it worked to higher orders in $\alpha'$ the large $O(D,D)$ symmetry could prove very powerful in restricting the form of $\alpha'$-corrections and organizing them. Indeed, in a remarkable paper \cite{Marques:2015vua} Marqu\'es and N\'u\~nez showed that the $O(D,D)$ covariant formulation can capture the first order $\alpha'$-correction to the bosonic and heterotic string.\footnote{Previous works on $\alpha'$-corrections in DFT include \cite{Hohm:2013jaa,Bedoya:2014pma,Hohm:2014xsa,Coimbra:2014qaa,Lee:2015kba}.} The corrections come about through a Green-Schwarz like modification of the double Lorentz transformations. This correction in fact induces a whole tower of $\alpha'$-corrections, by requiring the closure of the corrected transformations, and the resulting terms have the right form to be able to account also for the $\alpha'^2$-corrections to the bosonic and heterotic string \cite{Baron:2018lve,Baron:2020xel}.\footnote{More precisely, the $R^2$-terms depend on two parameters, $a$ and $b$, where $b=0$ (or $a=0$) gives the heterotic $R^2$-terms and $a=b$ gives the bosonic ones. The induced $\alpha'^2$-terms have coefficients $a^2$, $b^2$ and $ab$ and if the $ab$-terms contain $R^3$ while the $a^2$ and $b^2$ ones don't, this would match the form of these corrections in the bosonic and heterotic case \cite{Metsaev:1986yb}. A detailed check of this remains to be done.} Recently these results have been used to find the order $\alpha'$ correction to generalizations of T-duality known as non-abelian and Poisson-Lie T-duality \cite{Borsato:2020wwk,Hassler:2020tvz,Codina:2020yma} and closely related integrable deformations of the string sigma model \cite{Borsato:2020bqo} (see also \cite{Hassler:2020wnp,Hassler:2020xyj}).

Given the success in describing the first $\alpha'$-corrections to the bosonic and heterotic string in an $O(D,D)$ covariant way, it is natural to ask whether it can work to even higher orders in $\alpha'$. This is the question we will address here. Unfortunately, though perhaps not too surprisingly, the answer appears to be no.

To attack this problem we set up a systematic procedure for constructing $O(D,D)$ invariants in the frame-like formulation of DFT. In this formulation $O(D,D)$ (and generalized diffeomorphism) invariance is manifest and the non-trivial problem is to construct higher-derivative terms which are invariant under (corrected) double Lorentz transformations. The key idea is to work order by order in fields, which simplifies the required calculations a lot. The simplest invariant to construct is the two-derivative action itself. At the next order in $\alpha'$ we have an invariant which starts as the Riemann tensor squared. We show that constructing this invariant, we find precisely the $\alpha'$-correction to the $O(D,D)$ invariant Lagrangian found in \cite{Marques:2015vua}. Our procedure also uniquely fixes the required correction to the double Lorentz transformations, which was taken as input in \cite{Marques:2015vua}. This demonstrates the (expected) uniqueness of the $R^2$ correction (up to the free parameter interpolating between the heterotic and bosonic case). We then look for invariants involving higher powers of the Riemann tensor, namely $R^3$ and $R^4$. We show that there is no $R^3$ invariant. Such terms are known to appear in the bosonic string effective action at order $\alpha'^2$ \cite{Metsaev:1986yb}, but our findings are consistent with this since such terms will be generated as part of the $R^2$ invariant by requiring the corrected Lorentz transformations to close to higher orders in $\alpha'$ \cite{Baron:2020xel}.

Finally we come to the most interesting case of the $R^4$ invariant at order $\alpha'^3$ (four loops in $\sigma$-model perturbation theory). This is the first correction in the case of the type II string and takes the following form \cite{Gross:1986iv,Grisaru:1986vi,Freeman:1986zh}
\begin{equation}
S^{(3)}=\frac{\alpha'^3\zeta(3)}{3\cdot 2^{13}}\int d^{10}x\sqrt{-G}e^{-2\Phi}\left[(t_8t_8+\tfrac14\varepsilon_8\varepsilon_8)R^4+H\mbox{-terms}\right]\,,
\label{eq:alpha3correction}
\end{equation}
where $t_8t_8R^4$ denotes
\begin{equation}
t_8^{a_1\cdots a_8}t_8^{b_1\cdots b_8}R_{a_1a_2b_1b_2}R_{a_3a_4b_3b_4}R_{a_5a_6b_5b_6}R_{a_7a_8b_7b_8}
\end{equation}
and similarly for $\varepsilon_8\varepsilon_8R^4$. As customary we have defined $\varepsilon_8$ so that
\begin{equation}
\varepsilon_8^{a_1\cdots a_8}\varepsilon_8^{b_1\cdots b_8}=\tfrac12\varepsilon^{a_1\cdots a_8cd}\varepsilon^{b_1\cdots b_8}{}_{cd}
\end{equation}
and $t_8$ is defined by
\begin{equation}
t_{abcdefgh}M_1^{ab}M_2^{cd}M_3^{ef}M_4^{gh}
=
8\tr(M_1M_2M_3M_4)
-2\tr(M_1M_2)\tr(M_3M_4)
+\mathrm{cyclic}(234)
\label{eq:t8}
\end{equation}
for anti-symmetric matrices $M_{1,2,3,4}$. The same terms also occur (along with other terms) for the heterotic string \cite{Cai:1986sa,Gross:1986mw} and the bosonic string \cite{Jack:1988sw}. The fact that, unlike the lower corrections, it is proportional to the transcendental number $\zeta(3)$ means that it cannot be part of the lower invariants and must be independent. Interestingly we find that $O(D,D)$ invariance fixes precisely the correct $t_8t_8$-structure of the quartic Riemann terms, at the leading order in fields. However, when we analyze the conditions further, we find that it is not possible to complete this invariant at the fifth order in fields preserving global $O(D,D)$ and local Lorentz symmetry. Therefore, imposing $O(D,D)$ symmetry \emph{before} compactification, as is typically done in DFT, is not consistent with the string $\alpha'$-expansion at order $\alpha'^3$. Let us emphasize that our results do not imply that there is any problem with the duality symmetry in the compactified theory. Indeed, restricting to backgrounds with $d$ commuting isometries, it should be possible to express everything in an $O(d,d)$ covariant way, as was done at order $\alpha'$ in \cite{Meissner:1996sa} (see also \cite{Bergshoeff:1995cg,Eloy:2019hnl,Eloy:2020dko,Elgood:2020xwu,Ortin:2020xdm}). Indeed, this could be very useful to fix the form of higher $\alpha'$-corrections, for example T-duality on a circle has been recently argued to fix the form of the string effective action uniquely to order $\alpha'^3$ \cite{Garousi:2020mqn,Garousi:2020gio}.

The outline of the rest of the paper is as follows. After introducing what we need from the flux formulation of DFT in section \ref{sec:DFT} we analyze the problem of constructing $O(D,D)$ invariants at the leading order in the number of fields in section \ref{sec:leading}. We then go to the subleading order and, after showing how the two-derivative action is recovered, we formulate necessary conditions for the higher leading invariants to extend to the subleading order in fields. In section \ref{sec:R2} we analyze the $R^2$ invariant and show that we recover the first $\alpha'$ correction of \cite{Marques:2015vua} including the required correction to the Lorentz transformations. Next we show that there is no (independent) $R^3$ invariant, while in section \ref{sec:R4} we analyze the case of $R^4$ and show that while consistency requires it to take the expected form involving the tensor $t_8$ further analysis reveals an obstruction to completing it with terms of fifth order in fields. We end with some conclusions.

\section{\boldmath Elements of the \texorpdfstring{$O(D,D)$}{O(D,D)} covariant formulation}\label{sec:DFT}
Here we will introduce the elements of the $O(D,D)$ covariant formulation of DFT, which we will need. We will use the frame-like formulation of DFT \cite{Siegel:1993th,Siegel:1993xq,Hohm:2010xe}, where the structure group consists of two copies of the Lorentz group. More specifically we use the so-called flux formulation of \cite{Geissbuhler:2013uka,Marques:2015vua}. Throughout we assume that the section condition 
\begin{equation}
\partial_MY\partial^MZ=0
\end{equation}
for any $Y,Z$, is solved in the standard way, taking $\partial_M=(0,\partial_m)$, so that we are really just working with a rewriting of (super)gravity\footnote{The formalism we are using can describe the bosonic low-energy string effective action or (the bosonic sector of) the heterotic string effective action with the gauge fields set to zero. There exists a simple extension to include Ramond-Ramond fields (and fermions) and describe also type II strings, e.g. \cite{Hohm:2011dv,Coimbra:2011nw,Jeon:2012kd}.} as in generalized geometry.

The basic building block is the generalized (inverse) vielbein parametrized as
\begin{equation}
E_A{}^M=
\frac{1}{\sqrt2}
\left(
\begin{array}{cc}
e^{(+)a}{}_m-e^{(+)an}B_{nm} & e^{(+)am}\\
-e^{(-)}_{am}-e^{(-)}_a{}^nB_{nm} & e^{(-)}_a{}^m
\end{array}
\right)\,.
\label{eq:E}
\end{equation}
It transforms under $O(D,D)$ as $E_A{}^M(X)\rightarrow E_A{}^N(XR)R_N{}^M$, with $R_N{}^M$ a constant $O(D,D)$ matrix, and under double Lorentz transformations as $E_A{}^M\rightarrow\Lambda_A{}^BE_B{}^M$ with $\Lambda$ non-constant and block-diagonal $\Lambda=(\Lambda^{(+)},\,\Lambda^{(-)})$. The two sets of vielbeins $e^{(\pm)}$ for the metric $G_{mn}$ transform independently as $\Lambda^{(\pm)}e^{(\pm)}$ under the two Lorentz-group factors. To go to the standard supergravity picture one fixes the gauge $e^{(+)}=e^{(-)}=e$, leaving only one copy of the Lorentz-group. The dilaton $\Phi$ is encoded in the generalized dilaton $d$ defined as
\begin{equation}
e^{-2d}=e^{-2\Phi}\sqrt{-G}\,.
\end{equation}
There are two constant metrics, the $O(D,D)$ metric $\eta^{AB}$ and the generalized metric $\mathcal H^{AB}$, which take the form
\begin{equation}
\eta^{AB}=\eta_{AB}=
\left(
\begin{array}{cc}
	\hat\eta & 0\\
	0 & -\hat\eta
\end{array}
\right)\,,\qquad
\mathcal H^{AB}=
\left(
\begin{array}{cc}
	\hat\eta & 0\\
	0 & \hat\eta
\end{array}
\right)\,,
\end{equation}
where $\hat\eta=(-1,1,\ldots,1)$ is the $D$-dimensional Minkowski metric. From these we build the projection operators
\begin{equation}
P_\pm^{AB}=\frac12\left(\eta^{AB}\pm\mathcal H^{AB}\right)\,.
\label{eq:Ppm}
\end{equation}
The flat tangent space indices $A,B,\ldots$ are raised(lowered) with $\eta^{AB}$($\eta_{AB}$), while the generalized vielbein is used to convert between these indices and coordinate indices $M,N,\ldots$.\footnote{In particular we have the usual expressions for the $O(D,D)$ metric and the generalized metric in a coordinate basis
$$
\eta^{MN}=E_A{}^M\eta^{AB}E_B{}^N=
\left(
\begin{array}{cc}
	0 & 1\\
	1 & 0
\end{array}
\right)\,, \qquad
\mathcal H^{MN}=E_A{}^M\mathcal H^{AB}E_B{}^N=
\left(
\begin{array}{cc}
	G-BG^{-1}B & BG^{-1}\\
	-G^{-1}B & G^{-1}
\end{array}
\right)\,.
$$
}
We define the derivative with a flat index as
\begin{equation}
\partial_A=E_A{}^M\partial_M\,,\qquad \partial_M=(0,\partial_m)\,.
\end{equation}

The diffeomorphism and B-field gauge transformation invariant information in the generalized vielbein is contained in the generalized fluxes defined as
\begin{equation}
F_{ABC}=3\partial_{[A}E_B{}^ME_{C]M}\,,\qquad F_A=\partial^BE_B{}^ME_{AM}+2\partial_Ad\,.
\label{eq:fluxes}
\end{equation}
They are manifestly $O(D,D)$ invariant and they are the basic building blocks from which to construct an $O(D,D)$ invariant action. The generalized fluxes satisfy the following Bianchi identities
\begin{equation}
4\partial_{[A}F_{BCD]}=3F_{[AB}{}^EF_{CD]E}\,,\qquad
2\partial_{[A}F_{B]}=-(\partial^C-F^C)F_{ABC}\,.
\label{eq:Bianchi}
\end{equation}
We also have
\begin{equation}
[\partial_A,\partial_B]=F_{ABC}\,\partial^C\,.
\label{eq:d-comm}
\end{equation}

Using the projection operators (\ref{eq:Ppm}) we can split the generalized fluxes into different components. We will use a notation where we denote indices projected with $P_+$($P_-$) by over(under)lining them. The components of the generalized fluxes are
\begin{equation}
F_{\o a}\,,\quad F_{\u a}\,,\qquad F_{\o{abc}}\,,\quad F_{\u a\o{bc}}\,,\quad F_{\o a\u{bc}}\,,\quad F_{\u{abc}}\,.
\end{equation}
We will also do the same for the derivatives, e.g. $\partial_{\o a}=(P_+E)_{\o a}{}^m\partial_m$. In this formulation $O(D,D)$ invariance is manifest and so is diffeomorphism and B-field gauge transformation invariance. The price we pay for this is that local (double) Lorentz invariance is far from manifest. In fact the generalized fluxes transform similarly to connections as
\begin{equation}
\delta F_{ABC}=3\partial_{[A}\lambda_{BC]}+3\lambda_{[A}{}^DF_{BC]D}\,,\qquad
\delta F_A=\partial^B\lambda_{BA}+\lambda_A{}^BF_B\,,
\label{eq:deltaF}
\end{equation}
for an infinitesimal transformation $\Lambda=1+\lambda$. Using the fact that the non-zero components of $\lambda$ are $\lambda^{(+)}_{\o{ab}}$ and $\lambda^{(-)}_{\u{ab}}$ the non-trivial transformations under $\lambda^{(+)}$ are
\begin{equation}
\delta F_{\o{abc}}=3\partial_{[\o a}\lambda_{\o{bc}]}+3\lambda_{[\o a}{}^{\o d}F_{\o{bc}]\o d}\,,\quad
\delta F_{\u a\o{bc}}=\partial_{\u a}\lambda_{\o{bc}}+\lambda_{\u a}{}^{\u d}F_{\u d\o{bc}}+2\lambda_{[\o b}{}^{\o d}F_{|\u a\o d|\o c]}\,,\quad
\delta F_{\o a}=\partial^{\o b}\lambda_{\o{ba}}+\lambda_{\o a}{}^{\o b}F_{\o b}\,,
\label{eq:deltaF+}
\end{equation}
and under $\lambda^{(-)}$ the same with over-/underlined indices exchanged
\begin{equation}
\delta F_{\u{abc}}=3\partial_{[\u a}\lambda_{\u{bc}]}+3\lambda_{[\u a}{}^{\u d}F_{\u{bc}]\u d}\,,\quad
\delta F_{\o a\u{bc}}=\partial_{\o a}\lambda_{\u{bc}}+\lambda_{\o a}{}^{\o d}F_{\o d\u{bc}}+2\lambda_{[\u b}{}^{\u d}F_{|\o a\u d|\u c]}\,,\quad
\delta F_{\u a}=\partial^{\u b}\lambda_{\u{ba}}+\lambda_{\u a}{}^{\u b}F_{\u b}\,.
\label{eq:deltaF-}
\end{equation}
To construct $O(D,D)$ invariants, we need to find combinations of these six fields and their derivatives, which are invariant under the above transformations.\footnote{A direct argument that generalized diffeomorphism invariance requires the Lagrangian to be expressed in terms of the generalized fluxes, using the tools introduced in the next section, is provided in appendix \ref{app:L-F}.} Normally we would solve this problem by constructing covariant field strengths. However, in the present case such quantities don't exist. In particular, to leading order in fields, the would-be field strength of $F_{\o{abc}}$, i.e. $4\partial_{[\o a}F_{\o{bcd}]}$, vanishes by the Bianchi identities (\ref{eq:Bianchi}). These also imply that the would-be field strength of $F_{\o a}$ also does not exist. Finally, for $F_{\o a\u{bc}}$, which is similar to the spin connection, we can define the "curvature" \cite{Hronek:2020skb}\footnote{Exchanging the over(under)lined indices leads to the same object up to a sign due to the Bianchi identities.}
\begin{equation}
R_{\o{ab}\u{cd}}=2\partial_{[\o a}F_{\o b]\u{cd}}
-F_{\o{abe}}F^{\o e}{}_{\u{cd}}
-2F_{[\o a|\u c|}{}^{\u e}F_{\o b]\u{ed}}\,,
\label{eq:R}
\end{equation}
which transforms as
\begin{equation}
\delta R_{\o{ab}\u{cd}}=
2\lambda_{[\o a}{}^{\o e}R_{|\o e|\o b]\u{cd}}
+2\lambda_{[\u c}{}^{\u e}R_{|\o{ab}\u e|\u d]}
-F_{\o e\u{cd}}\partial^{\o e}\lambda_{\o{ab}}
+F_{\u e\o{ab}}\partial^{\u e}\lambda_{\u{cd}}\,.
\label{eq:deltaR}
\end{equation}
This is the closest we can come to a Riemann tensor, however the last two terms in the transformation show that it is not a Lorentz covariant object. Note however that at the leading order in fields it does behave like the Riemann tensor, $\delta R_{\o{ab}\u{cd}}\sim 0$, a fact that will be important later. Before proceeding to the construction of invariants, we will briefly describe the linearization of the generalized vielbein.

\subsection{Linearized level}
At some points in our calculations it will be convenient to linearize around flat space. Taking
\begin{equation}
G_{mn}=\eta_{mn}+h_{mn}\,,\qquad B_{mn}=b_{mn}\,,
\end{equation}
where $h_{mn}$ and $b_{mn}$ are the linear fluctuations of the metric and $B$-field, we find
\begin{equation}
E_A{}^M=E_A^{(0)M}+\hat e_A{}^M
\end{equation}
where
\begin{equation}
E_A^{(0)M}=\frac{1}{\sqrt2}(\delta_A^M+\eta_{AB}\delta^B_N\eta^{NM})\,,\qquad
\hat e_A{}^M=
\frac{1}{\sqrt2}\left(
\begin{array}{cc}
\tfrac12h^a{}_m-b^a{}_m & -\tfrac12h^{am}\\
-\tfrac12h_{am}-\lambda_{am}-b_{am} & -\tfrac12h_a{}^m+\lambda_a{}^m
\end{array}
\right)
\end{equation}
where $\lambda_{ab}$ are the parameters of the infinitesimal Lorentz transformation relating the two vielbeins, $e^{(-)}\sim(1+\lambda)e^{(+)}\sim 1+\frac12h+\lambda$. It is convenient to define
\begin{equation}
\hat e_{AB}\equiv
\hat e_A{}^ME^{(0)}_{BM}
=
-\frac12\left(
\begin{array}{cc}
b^{ab} & -h^a{}_b+b^a{}_b\\
h_a{}^b+b_a{}^b & b_{ab}+2\lambda_{ab}
\end{array}
\right)\,.
\end{equation}
Note that it is anti-symmetric, $\hat e_{AB}=\hat e_{[AB]}$. The linearization of the three-index generalized flux $F_{ABC}$ in (\ref{eq:fluxes}) becomes
\begin{equation}
F_{ABC}=3\partial_{[A}\hat e_{BC]}\,.
\label{eq:F-lin}
\end{equation}

\section{Invariants at leading order in fields}\label{sec:leading}
We wish to find $O(D,D)$ invariant Lagrangians and to simplify this task we will work order by order in fields. We will follow a similar approach to the classic paper by Utiyama \cite{Utiyama:1956sy}, which analyzed the possible gauge invariant Lagrangians in some standard gauge theories.

We assume that the action takes the form
\begin{equation}
S_n=\int dX\,e^{-2d}L_n\,,
\end{equation}
where $n$ denotes the order in fields and $L_n$ is constructed from the generalized fluxes and their flat derivatives. We will also assume that at leading order in fields the Lagrangian $L_n$ is invariant under double Lorentz transformations. We will therefore not consider Chern-Simons terms (at leading order in fields), since they are not very relevant for the questions we want to ask.

Let us focus on one of the Lorentz factors with parameters $\lambda^{(+)}$. To leading order in fields the non-trivial Lorentz transformations (\ref{eq:deltaF+}) become
\begin{equation}
\delta F_{\o{abc}}\sim3\partial_{[\o a}\lambda_{\o{bc}]}\,,\qquad
\delta F_{\u c\o{ab}}\sim\partial_{\u c}\lambda_{\o{ab}}\,,\qquad
\delta F_{\o a}\sim\partial^{\o b}\lambda_{\o{ba}}\,.
\end{equation}
We denote the derivatives of the Lagrangian with respect to the fields and their derivatives as
\begin{equation}
G^{\o a,A_1\cdots A_k}=\frac{\partial L_n}{\partial(\partial_{A_1\cdots A_k}F_{\o a})}\,,\quad
G^{\o{abc},A_1\cdots A_k}=\frac{\partial L_n}{\partial(\partial_{A_1\cdots A_k}F_{\o{abc}})}\,,\quad
G^{\o{ab}\u c,A_1\cdots A_k}=\frac{\partial L_n}{\partial(\partial_{A_1\cdots A_k}F_{\u c\o{ab}})}\,.
\label{eq:Gs}
\end{equation}
To leading order in fields the condition that the Lagrangian be invariant under the $\lambda^{(+)}$ transformations then reads
\begin{equation}
\sum_{k=0}^N
\left(
G^{\o a,A_1\cdots A_k}\partial_{A_1\cdots A_k}{}^{\o b}\lambda_{\o{ba}}
+3G^{\o{abc},A_1\cdots A_k}\partial_{A_1\cdots A_k\o c}\lambda_{\o{ab}}
+G^{\o{ab}\u c,A_1\cdots A_k}\partial_{A_1\cdots A_k\u c}\lambda_{\o{ab}}
\right)=0\,,
\end{equation}
where $N$ is the highest number of derivatives that occurs. We now follow the same approach as Utiyama. First we note that terms with different numbers of derivatives acting on $\lambda$ are clearly independent, so each term in the sum must vanish separately giving the conditions
\begin{equation}
G^{\o a,A_1\cdots A_k}\partial_{A_1\cdots A_k}{}^{\o b}\lambda_{\o{ba}}
+3G^{\o{abc},A_1\cdots A_k}\partial_{A_1\cdots A_k\o c}\lambda_{\o{ab}}
+G^{\o{ab}\u c,A_1\cdots A_k}\partial_{A_1\cdots A_k\u c}\lambda_{\o{ab}}
=0\qquad\forall k\,.
\label{eq:L-inv}
\end{equation}
Taking $k=0$ this becomes
\begin{equation}
G^{\o a}\partial^{\o b}\lambda_{\o{ba}}
+3G^{\o{abc}}\partial_{\o c}\lambda_{\o{ab}}
+G^{\o{ab}\u c}\partial_{\u c}\lambda_{\o{ab}}
=0\,.
\end{equation}
We have to be careful because, due to the section condition $\partial_{\o a}Y\partial^{\o a}Z+\partial_{\u a}Y\partial^{\u a}Z=0$, the last term can mix with the other two terms. This can of course only happen for terms in $G^{\o{ab}\u c}$ where the index $\u c$ is sitting on a derivative. To take this ambiguity into account, we let $H^{\o{ab}\u c}$ denote terms with the index $\u c$ sitting on a derivative and $H^{\o{abc}}$ be the same with the index $\u c$ replaced by $\o c$. The condition then splits into the two conditions
\begin{equation}
G^{\o{ab}\u c}+H^{\o{ab}\u c}=0\,,\qquad
\eta^{\o c[\o a}G^{\o b]}+3G^{\o{abc}}+H^{\o{abc}}=0\,.
\end{equation}
Since $H^{\o{abc}}$ comes from $H^{\o{ab}\u c}$ it cannot contain $\eta^{\o{ca}}$ so it cannot mix with the $\eta^{\o{ca}}G^{\o b}$-term. The same is true for $G^{\o{abc}}$ since it is anti-symmetric in all three indices. Therefore the last equation implies $G^{\o a}=0$ and $3G^{\o{abc}}+H^{\o{abc}}=0$. The last condition implies that all components of $H^{\o{abc}}$ vanish except the piece which is anti-symmetric in $\o{abc}$, but since $\o c$ is sitting on a derivative by assumption this means that also $\o a$ and $\o b$ are sitting on derivatives, i.e. it takes the form $H^{\o{abc}}=\partial^{[\o a}W\partial^{\o b}X\partial^{\o c]}YZ$ for some $W,X,Y,Z$. Clearly the derivatives cannot act on the same field since that would give zero due to anti-symmetry since the derivatives commute to leading order. Therefore $H$ has to contain at least three fields. Therefore, we have found that the Lagrangian can contain fields without a derivative only in the combination
\begin{equation}
F_{\o{abc}}\partial^{\o a}W\partial^{\o b}X\partial^{\o c}YZ
+3F_{\u a\o{bc}}\partial^{[\u a}W\partial^{\o b}X\partial^{\o c]}YZ\,.
\label{eq:d3-inv}
\end{equation}
Clearly by looking also at the $\lambda^{(-)}$ variation we will find the same but with over(under)lined indices exchanged, however this is not independent due to the section condition which implies that $F_{ABC}\partial^AX\partial^BY\partial^CZ=0$.

Next we consider the condition (\ref{eq:L-inv}) at $k=1$ which reads
\begin{equation}
G^{\o a,D}\partial_D{}^{\o b}\lambda_{\o{ba}}
+3G^{\o{abc},D}\partial_{D\o c}\lambda_{\o{ab}}
+G^{\o{ab}\u c,D}\partial_{D\u c}\lambda_{\o{ab}}
=0\,.
\end{equation}
Next we must split the index $D=(\o d,\u d)$. Note that we may assume that if this index is sitting on a derivative it takes only the value $\o d$, since we can always use the section condition in the original Lagrangian to arrange this. The last term can mix with the others in the same way as for $k=0$ but this just gives rise to the same invariant decorated by an extra derivative so we may ignore this for the moment. Then looking at the $\partial_{\u{dc}}\lambda_{\o{ab}}$-term we see that it can only vanish if $G^{\o{ab}\u c,\u d}$ is actually anti-symmetric in the two underlined indices. Similarly the $\partial_{\o{dc}}\lambda_{\o{ab}}$-terms imply that $G^{\o a,\o d}\propto\eta^{\o{ad}}$ and that $G^{\o{abc},\o d}$ is completely anti-symmetric. The latter however gives a trivial contribution since a term $G^{\o{abc},\o d}\partial_{\o d}F_{\o{abc}}$ in the Lagrangian then vanishes due to the Bianchi identity for $F_{\o{abc}}$. Therefore we may set $G^{\o{abc},\o d}=0$. We are left with the following condition
\begin{equation}
\left(\eta^{\o{ca}}G^{\o b,\u d}+3G^{\o{abc},\u d}+G^{\o{ab}\u d,\o c}\right)\partial_{\o c\u d}\lambda_{\o{ab}}=0\,.
\end{equation}
This requires $3G^{\o{abc},\u d}+G^{[\o{ab}|\u d|,\o c]}=0$ but this again leads to something trivial due to the Bianchi identities, namely $\partial_{[\o a}F_{|\u d|\o{bc}]}\sim\frac13\partial_{\u d}F_{\o{abc}}$. We conclude that $G^{\o{abc},\u d}=0$ and the remaining terms imply $G^{\o{ab}\u d,\o c}=\eta^{\o c[\o b}G^{\o a],\u d}$. Putting all this together we have found that the Lagrangian can contain $\partial F$ only in the combinations
\begin{align}
	(a)&\quad\partial_{[\o a}F_{\o b]\u{cd}}\nonumber\\
	(b)&\quad\partial^{\o a}F_{\o a}\nonumber\\
	(c)&\quad\partial_{\u a}F_{\o b}+\partial^{\o c}F_{\u a\o{bc}}\nonumber
\end{align}
or as (\ref{eq:d3-inv}) decorated by extra derivatives. From the $\lambda^{(-)}$ variation we find the same with over(under)lined indices exchanged, but these are again not independent.

Considering terms with more derivatives on the fields, corresponding to $k>1$ in (\ref{eq:L-inv}), does not lead to anything new. One just finds that the Lagrangian can depend also on derivatives of the objects found for $k=0$ and $k=1$.

\section{Invariants at subleading order}
We will now formulate the conditions for the leading order invariants to extend to the next order in fields. At this order, we can no longer require the Lagrangian to be Lorentz invariant, but we must allow it to transform by a total derivative.

\subsection{Two-derivative action}
Let us first consider the simplest case of the (two derivative) action itself. There is a unique leading order invariant at dimension 2 which does not have free indices namely $\partial^{\o a}F_{\o a}$. To extend this invariant to the next order in fields we take\footnote{While this is a total derivative at the leading order in fields this will not be the case at the next order.}
\begin{equation}
L=4\partial^{\o a}F_{\o a}+L_2,
\end{equation}
where $L_2$ denotes terms that are quadratic in fields. Requiring this to be invariant, up to a total derivative, under the $\lambda^{(+)}$ variation (\ref{eq:deltaF+}) gives the condition
\begin{equation}
2F^{\o{ab}C}\partial_C\lambda_{\o{ba}}
+4F^{\o b}\partial^{\o a}\lambda_{\o{ab}}
+G^{\o a}\partial^{\o b}\lambda_{\o{ba}}
+G^{\o{ab}\u c}\partial_{\u c}\lambda_{\o{ab}}
+3G^{\o{abc}}\partial_{\o c}\lambda_{\o{ab}}
+\partial^CT_C=0\,,
\end{equation}
where the $G$'s are defined as in (\ref{eq:Gs}) and $T_C$ encodes the total derivative terms. In this case it is easy to see that the total derivative terms are not needed and a solution is given by
\begin{equation}
T_C=0\,,\qquad
G^{\o a}=-4F^{\o a}\,,\quad
G^{\o{abc}}=\tfrac23F^{\o{abc}}\,,\quad
G^{\o{ab}\u c}=2F^{\u c\o{ab}}\,,
\end{equation}
which, upon integration, produces precisely the known Lagrangian  \cite{Geissbuhler:2013uka} (up to the overall sign and exchange of underlined and overlined indices)
\begin{equation}
L=4\partial^{\o a}F_{\o a}-2F^{\o a}F_{\o a}+F^{\u a\o{bc}}F_{\u a\o{bc}}+\tfrac13F^{\o{abc}}F_{\o{abc}}\,.
\label{eq:L2}
\end{equation}

\subsection{Higher invariants}
We are interested in possible higher order (in $\alpha'$) invariants that can be added to this lowest order action. We know from our analysis in the previous section that at the leading order in fields they must be constructed from the following combinations of fields (or derivatives of these)
\begin{align}
(a)&\quad F_{\o{abc}}\partial^{\o a}W\partial^{\o b}X\partial^{\o c}YZ+3F_{\u a\o{bc}}\partial^{[\u a}W\partial^{\o b}X\partial^{\o c]}YZ\nonumber\\
(b)&\quad R_{\o{ab}\u{cd}}\sim2\partial_{[\o a}F_{\o b]\u{cd}}\nonumber\\
(c)&\quad R\sim\partial^{\o a}F_{\o a}\nonumber\\
(d)&\quad R_{\u a\o b}\sim\partial_{\u a}F_{\o b}+\partial^{\o c}F_{\u a\o{bc}}\nonumber
\end{align}
However, (c) and (d) are, up to subleading terms, the generalized Ricci scalar and generalized Ricci tensor, which vanish by the equations of motion of the lowest order action (\ref{eq:L2}). Therefore any invariant built using these can be removed by field redefinitions. Furthermore (a) cannot lead to any invariants of dimension less than 10. This is because the derivatives must act on fields which are again of the forms listed above. The possibility of lowest dimension is of the form
\begin{equation}
F_{\o{abc}}\partial^{\o a}R\partial^{\o b}R\partial^{\o c}R+3F_{\u a\o{bc}}\partial^{[\u a}R\partial^{\o b}R\partial^{\o c]}R\,,
\end{equation}
with all the indices on the $R$'s contracted. This has dimension 10 and so occurs at order $\alpha'^4$. Therefore, since we will confine ourselves to invariants up to order $\alpha'^3$, they must be constructed out of $R_{\o{ab}\u{cd}}$ and its derivatives at the leading order in fields. The possible invariants are therefore $R^n$ for $n=2,3,4$ or $\partial^2R^2$, $\partial^4R^2$ or $\partial^2R^3$. However, that latter three can actually be removed by field redefinitions (at leading order in fields). To see this we note that the first of these has the form
\begin{equation}
R^{\o{ab}\u{cd}}\partial^{\o e}\partial_{\o e}R_{\o{ab}\u{cd}}\,.
\end{equation}
But we may use the fact that $\partial_{[\o e}R_{\o{ab}]\u{cd}}\sim 0$ to write this a something involving the divergence of $R$. But now we note that
\begin{equation}
\partial^{\o a}R_{\o{ab}\u{cd}}
\sim
2\partial^{\o a}\partial_{[\o a}F_{\o b]\u{cd}}
\sim
-2\partial^{\o a}\partial_{[\u c}F_{\u d]\o{ab}}
\sim
2\partial_{[\u c}\left(\partial_{\u d]}F_{\o b}+\partial^{\o a}F_{\u d]\o{ba}}\right)\,,
\end{equation}
where we used the Bianchi identities (\ref{eq:Bianchi}) in the second step. The final expression is proportional to the equations of motion (to leading order in fields). This shows that terms involving a divergence of $R$ can be removed by a field redefinition, modulo terms of higher order in fields. This rules out non-trivial $\partial^2R^2$ and $\partial^4R^2$ invariants. Finally the $\partial^2R^3$ terms have the three possible structures
\begin{equation}
R^{\o{ab}\u{ef}}\partial^{\o d}R_{\o{bc}\u f}{}^{\u g}\partial_{\o d}R^{\o c}{}_{\o a\u{ge}}\,,\qquad
R_{\o{ab}\u{ef}}\partial^{\o a}R^{\o{cd}\u{fg}}\partial^{\o b}R_{\o{cd}\u g}{}^{\u e}\,,\qquad
R^{\o{ab}\u{ef}}\partial_{\u e}R_{\o b}{}^{\o c\u{dg}}\partial_{\u f}R_{\o{ca}\u{dg}}\,.
\end{equation}
The first can be written (up to total derivatives) as terms of the form $RR\partial^2R$, which can be removed just as for $R\partial^2R$. The second and third involve (up to a total derivative) the divergence of $R$ and can therefore also be removed. This leaves us with $R^n$ for $n=2,3,4$ as the only possible invariants up to order $\alpha'^3$. Therefore our higher derivative corrections are of the form
\begin{equation}
L=R^n+L_{n+1}+L_{n+2}+\ldots\,,
\end{equation}
where the subscript on $L$ denotes the order in fields. We will analyze the conditions for $L_{n+1}$ to exist. 

Before we write these conditions, we must take one more complication into account. That is the fact that we can also correct the Lorentz transformations themselves at higher orders in $\alpha'$. In fact, this is needed to correctly account for the first $\alpha'$-correction for the bosonic and heterotic string. To allow for this we take the Lorentz transformation of the generalized vielbein to be
\begin{equation}
\delta E_A{}^ME_{BM}=\lambda_{AB}+\hat\lambda_{AB}\,,
\label{eq:deltaE}
\end{equation}
where $\hat\lambda_{AB}$ are higher order in $\alpha'$ and constructed out of the parameters $\lambda_{AB}$ and the fields. We may assume that the only non-zero components of $\hat\lambda_{AB}$ are $\hat\lambda_{\o a\u b}=-\hat\lambda_{\u b\o a}$ since the diagonal components can be absorbed into the gauge parameters $\lambda_{\o{ab}}$ and $\lambda_{\u{ab}}$.\footnote{We have to demand that these transformations close to the order we are working. We have
$$
[\delta,\delta']E_A{}^ME_{BM}
=
[\lambda',\lambda]_{AB}
+[\lambda',\hat\lambda]_{AB}
-[\lambda,\hat\lambda']_{AB}
+\delta\hat\lambda'_{AB}
-\delta'\hat\lambda_{AB}
+\mbox{higher order terms}
\,.
$$
Closure of the transformations requires the RHS to be again of the form (\ref{eq:deltaE}) plus a generalized diffeomorphism
$$
\delta E_A{}^ME_{BM}=2\partial_{[A}Y_{B]}-F_{ABC}Y^C\,.
$$
}

The higher order correction to the Lorentz transformations will lead to a mixing of terms from the lowest order Lagrangian and the higher $\alpha'$-terms in the Lagrangian. The variation of the lowest order Lagrangian (\ref{eq:L2}) gives rise to
\begin{equation}
4(\partial^{\o a}-F^{\o a})\partial^{\u b}\hat\lambda_{\u b\o a}
+4\partial^{\o a}(\hat\lambda_{\o a\u b}F^{\u b})
+4\hat\lambda_{\o a\u b}\partial^{\u b}F^{\o a}
+4F^{\u a\o{bc}}\partial_{\o b}\hat\lambda_{\o c\u a}\,,
\end{equation}
at leading order in fields. However, the first two terms lead to total derivative terms and partially integrating also the last term we find
\begin{equation}
4(\partial^{\u a}F^{\o b}+\partial_{\o c}F^{\u a\o{bc}})\hat\lambda_{\o b\u a}+\mbox{total derivatives}\,.
\end{equation}

Now we can state the condition that a leading order invariant, of the form $R^n$, can be completed to the next order in fields to
\begin{equation}
L_n(R)+L_{n+1}+\ldots\,.
\end{equation}
The condition coming from (possibly corrected) Lorentz invariance is then, looking only at the $\lambda^{(+)}$-terms in (\ref{eq:deltaF+}) and (\ref{eq:deltaR}),\footnote{Note that the first two terms in $\delta R$ in (\ref{eq:deltaR}) don't contribute.}
\begin{align}
&{}-G^{\o{ab}\u{de}}F_{\o c\u{de}}\partial^{\o c}\lambda_{\o{ab}}
\nonumber\\
&{}
+\sum_{k=0}^N
\left(
G^{\o a,A_1\cdots A_k}\partial_{A_1\cdots A_k}{}^{\o b}\lambda_{\o{ba}}
+3G^{\o{abc},A_1\cdots A_k}\partial_{A_1\cdots A_k\o c}\lambda_{\o{ab}}
+G^{\o{ab}\u c,A_1\cdots A_k}\partial_{A_1\cdots A_k\u c}\lambda_{\o{ab}}
\right)
\nonumber\\
&{}
+4(\partial^{\u a}F^{\o b}+\partial_{\o c}F^{\u a\o{bc}})\hat\lambda_{\o b\u a}
=\partial_AT^A\,,
\label{eq:variation}
\end{align}
where we have defined
\begin{equation}
G^{\o{ab}\u{cd}}=\frac{\partial L_n}{\partial R_{\o{ab}\u{cd}}}\,,
\end{equation}
while the remaining $G$'s are define as in (\ref{eq:Gs}) with $L_n$ replaced by $L_{n+1}$. Here $T^A$ accounts for possible total derivative terms. Expanding $\hat\lambda$ and $T^A$ as\footnote{We may use the freedom to add a generalized diffeomorphism to the Lorentz transformation to set for example all $\hat\lambda_{\o c\u d}{}^{\o{ab},\o a_1\cdots\o a_k}=0$.}
\begin{equation}
T^A=\sum_{k=0}^NT^{\o{ab}A,A_1\cdots A_k}\partial_{A_1\cdots A_k}\lambda_{\o{ab}}\,,\qquad
\hat\lambda_{\o c\u d}=\sum_{k=0}^N\hat\lambda_{\o c\u d}{}^{\o{ab},A_1\cdots A_k}\partial_{A_1\cdots A_k}\lambda_{\o{ab}}\,,
\end{equation}
we get from the terms with no derivatives on $\lambda$ the condition
\begin{equation}
4(\partial^{\u a}F^{\o c}+\partial_{\o d}F^{\u a\o{cd}})\hat\lambda_{\o c\u a}{}^{\o{ab}}
=
\partial_{\o c}T^{\o{abc}}+\partial_{\u c}T^{\o{ab}\u c}\,,
\label{eq:cond-1}
\end{equation}
while the terms with one derivative on $\lambda$ give the conditions
\begin{align}
-G^{\o{ab}\u{de}}F^{\o c}{}_{\u{de}}
-G^{[\o a}\eta^{\o b]\o c}
+3G^{\o{abc}}
+H^{\o{abc}}
+4(\partial^{\u d}F^{\o e}+\partial_{\o f}F^{\u d\o{ef}})\hat\lambda_{\o e\u d}{}^{\o{ab},\o c}
=&
T^{\o{abc}}
+\partial_{\o d}T^{\o{abd},\o c}
+\partial_{\u d}T^{\o{ab}\u d,\o c}\,,
\label{eq:cond-2}
\\
G^{\o{ab}\u c}
+H^{\o{ab}\u c}
+4(\partial^{\u d}F^{\o e}+\partial_{\o f}F^{\u d\o{ef}})\hat\lambda_{\o e\u d}{}^{\o{ab},\u c}
=&
T^{\o{ab}\u c}
+\partial_{\o d}T^{\o{abd},\u c}
+\partial_{\u d}T^{\o{ab}\u d,\u c}\,.
\label{eq:cond-3}
\end{align}
There will of course be more conditions coming from the terms in the variation with two or more derivatives on $\lambda$ but the above conditions turn out to be sufficient for our purposes. Recall that here $H^{\o{ab}\u c}$ denotes (arbitrary) terms involving $\partial^{\u c}$ and $H^{\o{abc}}$ is the same with $\partial^{\u c}\rightarrow\partial^{\o c}$.

It will be convenient to have simpler conditions to deal with, which will be necessary but not sufficient. This is obtained by noting that the combination $\partial^{\u d}F^{\o e}+\partial_{\o f}F^{\u d\o{ef}}$ vanishes (to this order in fields) when the equations of motion are imposed. Therefore if we restrict ourselves to field configurations solving the equations of motion the $\hat\lambda$ contributions drop out. In fact it will be simpler to restrict the field configurations even more to require $F_A=\partial^{\o a}F_{\o a BC}=\partial^{\u a}F_{\u a BC}=0$. Note that this is consistent with the Bianchi identities. The first condition now becomes 
\begin{equation}
\partial_{\o c}T^{\o{abc}}+\partial_{\u c}T^{\o{ab}\u c}\sim 0
\end{equation}
and we may set $T^{\o{ab}C}\sim0$. To see this we note that we may assume that there are no cancellations between the two terms due to the section condition, since such contributions would just cancel out in $\partial_AT^A$. Therefore the only way the equation can hold \emph{without} restricting the field configurations is if $T^{\o{ab}C}=\partial_DT^{\o{ab}[CD]}$. But then we can write $T^{\o{ab}C}\lambda_{\o{ab}}=\partial_D(T^{\o{ab}[CD]}\lambda_{\o{ab}})-T^{\o{ab}[CD]}\partial_D\lambda_{\o{ab}}$ and the first term drops out and the second has a derivative on $\lambda$ and such terms are already accounted for. Finally, suppose we have $T^{\o{ab}C}\nsim0$, e.g. $T^{\o{abc}}=F^{\u e\o{cd}}W^{\o{ab}}{}_{\o d\u e}$ for some $W^{\o{ab}}{}_{\o d\u e}$. We must then have $F^{\u e\o{cd}}\partial_{\o c}W^{\o{ab}}{}_{\o d\u e}\sim0$ which requires $W^{\o{ab}}{}_{\o d\u e}=\partial_{\o d}W^{\o{ab}}{}_{\u e}$. But if this is the case we may again take the derivative to act on $\lambda$ instead (up to terms involving a divergence of $F^{\u e\o{cd}}$) and such terms are already accounted for. Therefore we may drop $T^{\o{ab}C}$ completely and the remaining conditions reduce to
\begin{equation}
-G^{\o{ab}\u{de}}F^{\o c}{}_{\u{de}}
-G^{[\o a}\eta^{\o b]\o c}
+3G^{\o{abc}}
+H^{\o{abc}}
\sim\,
\partial_DT^{\o{ab}D,\o c}\,,\qquad
G^{\o{ab}\u c}+H^{\o{ab}\u c}\sim\,\partial_DT^{\o{ab}D,\u c}\,.
\label{eq:cond2-3}
\end{equation}
These simplified conditions turn out to be enough to rule out $O(D,D)$ invariants of the form $R^3$ and $R^4$. An invariant of the form $R^2$ is known to exist and we will now see how it can be derived from these conditions.

\section{\boldmath \texorpdfstring{$R^2$}{R**2} invariant}\label{sec:R2}
The only possible structure of the form $R^2$ is
\begin{equation}
L_2=R^{\o{ab}\u{cd}}R_{\o{ab}\u{cd}}\,.
\end{equation}
We want to find $L_3$, cubic in fields, such that
\begin{equation}
L_2+L_3
\end{equation}
is Lorentz invariant up to terms of higher order in fields and total derivatives. We will show that $L_3$ exists and reproduces the known cubic terms in the $\alpha'$-corrected action. At the same time, our analysis fixes the correction to the Lorentz transformations again reproducing the known correction and showing that the result at this order is unique.

We will first work in the approximation $F_A\sim\partial^{\o a}F_{\o a BC}\sim\partial^{\u a}F_{\u a BC}\sim0$. Then we need to satisfy the conditions (\ref{eq:cond2-3}), which become,
\begin{equation}
-2R^{\o{ab}\u{de}}F^{\o c}{}_{\u{de}}
-G^{[\o a}\eta^{\o b]\o c}
+3G^{\o{abc}}
+H^{\o{abc}}
\sim\,
\mbox{total derivatives}\,,\qquad
G^{\o{ab}\u c}+H^{\o{ab}\u c}\sim\,\mbox{total derivatives}\,.
\end{equation}
If we symmetrize the first equation in $\o{bc}$ and drop terms where these indices are sitting on a derivative the last two terms on the LHS drop out and we find the condition
\begin{equation}
-2R^{\o a(\o b}{}_{\u{de}}F^{\o c)\u{de}}
-\tfrac12G^{\o a}\eta^{\o b\o c}
+\tfrac12G^{(\o b}\eta^{\o c)\o a}
\sim
\partial^{\o b,\o c}-\mbox{terms}+\mbox{total derivatives}\,.
\end{equation}
Furthermore, since
\begin{equation}
R^{\o a(\o b}{}_{\u{de}}F^{\o c)\u{de}}
\sim
\tfrac12\partial^{\o a}(F^{\o b}{}_{\u{de}}F^{\o c\u{de}})
-\partial^{(\o b}F^{|\o a|}{}_{\u{de}}F^{\o c)\u{de}}
\end{equation}
we find that $G^{\o a}\sim$ total derivatives. Going back to the original equations and noting that
\begin{equation}
R^{\o{ab}}{}_{\u{de}}F^{\o c\u{de}}
\sim
\tfrac32R^{[\o{ab}}{}_{\u{de}}F^{\o c]\u{de}}
+\partial^{[\o a}(F^{\o b]}{}_{\u{de}}F^{\o c\u{de}})
-\partial^{\o c}F^{[\o a}{}_{\u{de}}F^{\o b]\u{de}}
\end{equation}
we read off
\begin{align}
G^{\o{abc}}\sim&\, R^{[\o{ab}}{}_{\u{de}}F^{\o c]\u{de}}+\mbox{ total derivatives}\,,\nonumber\\
H^{\o{abc}}\sim&\,-2\partial^{\o c}F^{[\o a}{}_{\u{de}}F^{\o b]\u{de}}+\mbox{ total derivatives}\,,\label{eq:R2-approx}\\
G^{\o{ab}\u c}\sim&\,2\partial^{\u c}F^{[\o a}{}_{\u{de}}F^{\o b]\u{de}}+\mbox{ total derivatives}\,.\nonumber
\end{align}
Clearly we will find the same conditions with overlined and underlined indices exchanged by looking at the $\lambda^{(-)}$-variation. The fact that $G^{\o{abc}}$ depends on $F_{\o c\u{ab}}$ while $G^{\o{ab}\u c}$ does not depend on $F_{\o{abc}}$ is potentially in conflict with the integrability condition
\begin{equation}
\frac{\partial^2L}{\partial F_{\o d\u{ef}}\partial F_{\o{abc}}}=\frac{\partial^2L}{\partial F_{\o{abc}}\partial F_{\o d\u{ef}}}
\qquad\mbox{or}\qquad
\frac{\partial G^{\o{abc}}}{\partial F_{\o d\u{ef}}}=\frac{\partial G^{\u{ef}\o d}}{\partial F_{\o{abc}}}\,.
\end{equation}
However, in the present case this condition becomes
\begin{equation}
\eta^{\o d[\o c}R^{\o{ab}]}{}_{\u{ef}}\sim\mbox{ total derivatives}\,,
\end{equation}
which is indeed satisfied since $R^{\o{ab}}{}_{\u{ef}}\sim2\partial^{[\o a}F^{\o b]}{}_{\u{ef}}$. Therefore we do not encounter a problem in this case, but we will see that in the $R^4$ case we are not so lucky.

So far we worked in the approximation $F_A\sim\partial^{\o a}F_{\o a BC}\sim\partial^{\u a}F_{\u a BC}\sim0$. We must now go back and solve the general conditions (\ref{eq:cond-1}), (\ref{eq:cond-2}) and (\ref{eq:cond-3}) including these terms. The first step is to determine the total derivative terms $T^A$. Consider the first condition (\ref{eq:cond-1}). We will now argue that we may set $T^{\o{ab}C}=0$. Since we may assume that $T^{\o{ab}C}$ contains only terms with $F_A$ or a divergence of $F_{ABC}$ we find that in this case it can consist only of the following terms
\begin{align}
&F^{\o{ab}D}\partial^EF^C{}_{DE}\,,\quad
F^{CE[\o a}\partial_DF^{\o b]D}{}_E\,,\quad
F_D\partial^CF^{\o{ab}D}\,,\quad
F^{[\o a}\partial_DF^{\o b]CD}\,,\quad
F^C\partial_DF^{\o{ab}D}\,,\quad
\nonumber\\
&
\partial^{[\o a}F_DF^{\o b]CD}\,,\quad
\partial^CF_DF^{\o{ab}D}\,,\quad
F^{[\o a}\partial^{\o b]}F^C\,,
\end{align}
where each projection of the indices $CDE$ should be considered an independent term. The fact that $\partial_CT^{\o{ab}C}$ is proportional to $\partial^{\u a}F^{\o c}+\partial_{\o d}F^{\u a\o{cd}}$ implies that, in particular, there should not be terms with two derivatives acting on one field. This reduces the above possibilities to only
\begin{equation}
F^{\o{ab}E}\partial_DF^{[CD]}{}_E\,,\quad
F^{[\o a}\partial_DF^{\o b][CD]}\,.
\end{equation}
It is now easy to see that the only solution to (\ref{eq:cond-1}) is $T^{\o{ab}C}=\hat\lambda_{\o c\u d}{}^{\o{ab}}=0$. It remains only to solve the conditions (\ref{eq:cond-2}) and (\ref{eq:cond-3}) or
\begin{align}
-2R^{\o{ab}\u{de}}F^{\o c}{}_{\u{de}}
-G^{[\o a}\eta^{\o b]\o c}
+3G^{\o{abc}}
+H^{\o{abc}}
+4(\partial^{\u d}F^{\o e}+\partial_{\o f}F^{\u d\o{ef}})\hat\lambda_{\o e\u d}{}^{\o{ab},\o c}
=\,&
\partial_{\o d}T^{\o{abd},\o c}
+\partial_{\u d}T^{\o{ab}\u d,\o c}\,,
\label{eq:R2-1}
\\
G^{\o{ab}\u c}
+H^{\o{ab}\u c}
=\,&
\partial_{\o d}T^{\o{abd},\u c}
+\partial_{\u d}T^{\o{ab}\u d,\u c}\,.
\label{eq:R2-2}
\end{align}
Here we have used the freedom in making a generalized diffeomorphism to set $\hat\lambda_{\o e\u d}{}^{\o{ab},\u c}=0$. To solve these we note that $T^{\o{ab}C,D}\sim F^2$ and the possible terms are (each projection of the index $E$ is an independent term)
\begin{align}
&F^{\o{ab}E}F^{CD}{}_E\,,\quad
F^{CE[\o a}F^{\o b]D}{}_E\,,\quad
F^{[\o a}F^{\o b]CD}\,,\quad
F^CF^{D\o{ab}}\,,\quad
F^DF^{C\o{ab}}\,,\nonumber\\
&
\eta^{CD}U^{\o{ab}}\,,\quad
\eta^{C[\o a}V^{\o b]D}\,,\quad
\eta^{D[\o a}W^{\o b]C}\,,
\end{align}
for some $U,V,W$ quadratic in $F$'s. We will assume that the Lagrangian does not contain terms involving a divergence of $F$, which we can always arrange by partial integrations. This means that in the above conditions the terms involving divergences of $F$ can only be canceled by the $\hat\lambda$-terms. Since there are no such terms in the second equation it is not hard to see that the only solution is
\begin{equation}
T^{\o{ab}\u d,\u c}=
c_1F^{\o{abe}}F_{\o e}{}^{\u{cd}}
+c_2F_{\u e}{}^{\o{ab}}F^{\u{cde}}
+\eta^{\u{cd}}U_1^{\o{ab}}
\,,\qquad
T^{\o{abd},\u c}=
c_1F^{\o{abe}}F^{\u c\o d}{}_{\o e}
+c_2F_{\u e}{}^{\o{ab}}F^{\o d\u{ec}}
+\eta^{\o d[\o a}W_1^{\o b]\u c}\,,
\end{equation}
for some $c_1,c_2$ and
\begin{align}
G^{\o{ab}\u c}=&
c_1\partial^{\o d}F^{\o{abe}}F^{\u c}{}_{\o{de}}
+c_1\partial_{\u d}F^{\o{abe}}F_{\o e}{}^{\u{cd}}
+c_2\partial_{\o d}F_{\u e}{}^{\o{ab}}F^{\o d\u{ec}}
+c_2\partial_{\u d}F_{\u e}{}^{\o{ab}}F^{\u{cde}}
+2c_1F^{\o{abe}}\partial^{\u c}F_{\o e}
\nonumber\\
&{}
-2c_1F^{\o{ab}}{}_{\o e}\partial^{(\o e}F^{\u c)}
+2c_2F^{\u e\o{ab}}\partial^{\u c}F_{\u e}
-2c_2F_{\u e}{}^{\o{ab}}\partial^{(\u e}F^{\u c)}
+\partial^{[\o a}W_1^{\o b]\u c}
+\partial^{\u c}U_1^{\o{ab}}
-H^{\o{ab}\u c}\,.
\end{align}
Looking at the structure of the divergence terms in (\ref{eq:R2-1}) a little work shows that we must take
\begin{align}
T^{\o{abd},\o c}
=&
c_3F^{\o{ab}\u e}F_{\u e}{}^{\o{cd}}
+c_4F^{\u e\o d[\o a}F_{\u e}{}^{\o b]\o c}
+\eta^{\o{cd}}U_2^{\o{ab}}
+\eta^{\o d[\o a}V_2^{\o b]\o c}
+\eta^{\o c[\o a}W_2^{\o b]\o d}\,,
\\
T^{\o{ab}\u d,\o c}
=&
c_5F_{\u e}{}^{\o{ab}}F^{\o c\u{de}}
+c_6F^{[\o a|\u{de}|}F_{\u e}{}^{\o b]\o c}
+\eta^{\o c[\o a}W_3^{\o b]\u d}\,.
\end{align}
Using this in (\ref{eq:R2-1}) we find $c_3=c_4=0$, $W_3^{\o b\u d}=0$,
\begin{equation}
V_2^{\o{bc}}=
\tfrac12(c_5-4)F^{\o b\u{de}}F^{\o c}{}_{\u{de}}
+c_7F_{\u e}F^{\u e\o{bc}}
+\eta^{\o{bc}}V\,,\qquad
W_2^{\o{bd}}=c_8F^{\u e\o{fb}}F_{\u e\o f}{}^{\o d}
\end{equation}
and
\begin{align}
\hat\lambda_{\o e\u d}{}^{\o{ab},\o c}
=&
-\tfrac14c_5F_{\u d}{}^{\o{ab}}\delta_{\o e}^{\o c}%
-\tfrac14c_8F_{\u d\o e}{}^{[\o a}\eta^{\o b]\o c}\,,
\nonumber\\
G^{\o{abc}}
=&
-\tfrac12(c_5-4)\partial^{[\o a}F^{\o b}{}_{\u{de}}F^{\o c]\u{de}}%
+\tfrac12c_7\partial^{[\o a}F_{\u e}F^{|\u e|\o{bc}]}
+\tfrac13c_7F^{\u e}\partial_{\u e}F^{\o{abc}}\,,
\\
H^{\o{abc}}
=&
\partial^{\o c}U_2^{\o{ab}}
+\tfrac12(c_5-4)\partial^{\o c}F^{[\o a}{}_{\u{de}}F^{\o b]\u{de}}%
+\tfrac12(2c_5-c_7)\partial^{\o c}F_{\u e}F^{\u e\o{ab}}%
-c_7F_{\u e}\partial^{\o c}F^{\u e\o{ab}}\,,
\nonumber\\
G^{\o a}
=&
c_8\partial^{\o d}F^{\u e\o{fa}}F_{\u e\o{fd}}
-c_8\partial_{\u d}F_{\o e}F^{\u d\o{ea}}
-\partial^{\o a}V\,.
\nonumber
\end{align}
Plugging $H^{\o{ab}\u c}$ into the expression for $G^{\o{ab}\u c}$ and requiring that there be no terms involving $\partial_{[A}F_{B]}$, since these contain a divergence of $F$ by the Bianchi identities (\ref{eq:Bianchi}), gives $c_1=0$ and $c_7=2c_5-4c_2$.

Finally, to complete the analysis we need to look also at the terms in the variation of the Lagrangian with two derivatives on $\lambda$ (we cannot have more than two derivatives at this order in $\alpha'$). From (\ref{eq:variation}) we get the conditions (note that $\hat\lambda_{\o e\u d}{}^{\o{ab},CD}=0$, $T^{\o{ab}C,DE}=0$ and $H^{\o{ab}\u c,D}=0$ since there aren't enough derivatives available at this order in $\alpha'$)
\begin{align}
-G^{[\o a,|(\o d}\eta^{\o c)|\o b]}
+3G^{\o{ab}(\o c,\o d)}
=&
T^{\o{ab}(\o c,\o d)}\,,
\\
-G^{[\o a,|\u d|}\eta^{\o b]\o c}
+3G^{\o{abc},\u d}
+G^{\o{ab}\u d,\o c}
=&
T^{\o{abc},\u d}
+T^{\o{ab}\u d,\o c}\,,
\\
G^{\o{ab}(\u c,\u d)}
=&T^{\o{ab}(\u c,\u d)}\,.
\end{align}
We may use Bianchi identities to set $G^{\o{abc},\u d}=0$. Plugging in the $T$'s we get (note that for example $G^{\o{ab}(\u c,\u d)}$ cannot have a term proportional to $\eta^{\u{cd}}$ since this would mean divergence terms in the Lagrangian contrary to our assumptions)
\begin{equation}
U_1^{\o{ab}}=U_2^{\o{ab}}=0\,,\qquad V=0
\end{equation}
and
\begin{align}
G^{\o{ab}(\o c,\o d)}=&G^{\o{ab}(\u c,\u d)}=0\,, & G^{\o a,\o c}=&(c_2-2)F^{\o a\u{de}}F^{\o c}{}_{\u{de}}+2(c_5-2c_2)F_{\u e}F^{\u e\o{ac}}+c_8F^{\u d\o{ea}}F_{\u d\o e}{}^{\o c}\,,
\nonumber\\
G^{\o a,\u d}=&W_1^{\o a\u d}\,, & G^{\o{ab}\u d,\o c}=&(c_5-c_2)F_{\u e}{}^{\o{ab}}F^{\o c\u{de}}+c_6F^{[\o a|\u{de}|}F_{\u e}{}^{\o b]\o c}\,.
\end{align}
Finally, imposing the integrability conditions, we find $c_5=2c_2$, $c_6=c_8=0$, $W_1^{\o a\u d}=0$,
\begin{equation}
G^{\o a}=G^{\o a,\u d}=G^{\o{ab}(\o c,\o d)}=G^{\o{ab}(\u c,\u d)}=0
\end{equation}
and we have the one-parameter family of solutions\footnote{Note that the apparent difference compared to (\ref{eq:R2-approx}) is accounted for by the total derivative terms.}
\begin{align}
\hat\lambda_{\o e\u d}{}^{\o{ab},\o c}=&-\tfrac12aF_{\u d}{}^{\o{ab}}\delta_{\o e}^{\o c}\,,
\nonumber\\
G^{\o{abc}}=&(2-a)\partial^{[\o a}F^{\o b}{}_{\u{de}}F^{\o c]\u{de}}
\nonumber\\
G^{\o{ab}\u c}=&
-2a\partial^{(\u d}F^{\u c)}F_{\u d}{}^{\o{ab}}
+a\partial_{\o d}F_{\u e}{}^{\o{ab}}F^{\o d\u{ec}}
+a\partial_{\u d}F_{\u e}{}^{\o{ab}}F^{\u{cde}}
-(a-2)\partial^{\u c}F^{[\o a}{}_{\u{de}}F^{\o b]\u{de}}\,,
\\
G^{\o a,\o c}=&(a-2)F^{\o a\u{de}}F^{\o c}{}_{\u{de}}\,,
\nonumber\\
G^{\o{ab}\u d,\o c}=&aF_{\u e}{}^{\o{ab}}F^{\o c\u{de}}\,,\nonumber
\end{align}
where we renamed $c_2$ to $a$. Remembering that the $\lambda^{(-)}$ variation will give us similar expressions where the over(under)lined indices are interchanged we finally find the Lagrangian
\begin{align}
L=&R^{\o{ab}\u{cd}}R_{\o{ab}\u{cd}}
-a\partial_{\u e}F^{\u c}F^{\u e\o{ab}}F_{\u c\o{ab}}
+aF^{\u{abc}}F_{\u a}{}^{\o{de}}\partial_{\u b}F_{\u c\o{de}}
-aF^{\o c\u{de}}F_{\u d}{}^{\o{ab}}\partial_{\o c}F_{\u e\o{ab}}
\nonumber\\
&{}
-b\partial_{\o e}F^{\o c}F^{\o e\u{ab}}F_{\o c\u{ab}}
+bF^{\o{abc}}F_{\o a}{}^{\u{de}}\partial_{\o b}F_{\o c\u{de}}
-bF^{\u c\o{ab}}F_{\o a}{}^{\u{de}}\partial_{\u c}F_{\o b\u{de}}
+\mathcal O(F^4)\,,
\end{align}
with $a$ and $b$ satisfying $a+b=2$. Up to the overall coefficient and total derivative terms this action coincides with the two-parameter action constructed in \cite{Marques:2015vua}. This is easy to see using the simplified form given in \cite{Hronek:2020skb} (see also \cite{Baron:2017dvb}). Note that our derivation fixes the modification of the double Lorentz transformations (\ref{eq:deltaE}) at the same time to be
\begin{equation}
\delta E_{\o a}{}^ME_{\u bM}=-\frac{a}{2}\partial_{\o a}\lambda^{\o{cd}}F_{\u b\o{cd}}+\frac{b}{2}\partial_{\u b}\lambda^{\u{cd}}F_{\o a\u{cd}}\,,
\end{equation}
in agreement with the transformations proposed in \cite{Marques:2015vua}. Having reproduced the known results up to order $\alpha'$ we will now look for $O(D,D)$ invariants at order $\alpha'^2$ and $\alpha'^3$.

\section{\boldmath \texorpdfstring{$R^3$}{R**3} invariant}
There is only one possible structure for the $R^3$-terms namely
\begin{equation}
L_3=R^{\o a}{}_{\o b\u{de}}R^{\o{bc}\u{ef}}R_{\o{ca}\u f}{}^{\u d}\,.
\end{equation}
We will now show that it is not possible to find an $L_4$, quartic in the fields, such that $L_3+L_4$ is Lorentz invariant up to total derivatives and higher order terms in the fields. To do this we look at the necessary conditions (\ref{eq:cond2-3}) where we are neglecting terms involving $F_A$ or a divergence of $F_{ABC}$. If we symmetrize the first equation in $\o{bc}$ and drop terms where one of these indices is sitting on a derivative we find
\begin{equation}
3F^{(\o c}{}_{\u{de}}R^{\o b)}{}_{\o f\u g}{}^{\u e}R^{\o{af}\u{gd}}
+\tfrac12G^{\o a}\eta^{\o b\o c}
-\tfrac12G^{(\o b}\eta^{\o c)\o a}
\sim
\partial^{\o b,\o c}-\mbox{terms}
+\mbox{total derivatives}\,.
\end{equation}
This simplifies further to
\begin{equation}
3F^{(\o c}{}_{\u{de}}\partial_{\o f}F^{\o b)\u e}{}_{\u g}\partial^{\o a}F^{\o f\u{gd}}
+\tfrac12G^{\o a}\eta^{\o b\o c}
-\tfrac12G^{(\o b}\eta^{\o c)\o a}
\sim
\partial^{\o b,\o c}-\mbox{terms}
+\mbox{total derivatives}\,.
\end{equation}
Linearizing the first term using (\ref{eq:F-lin}) it becomes (up to the terms we are not keeping track of)
\begin{equation}
-6\hat e_{(\o b|\u d|}\partial^{\o f\u g}\hat e_{\o c)\u e}\partial_{\o a}{}^{\u{de}}\hat e_{\o f\u g}\,,
\end{equation}
which clearly cannot be written as a total derivative or canceled by the $G^{\o a}$-terms. This shows that no (independent) $O(D,D)$ invariant $R^3$-terms exist. Of course, $R^3$-terms can be, and are, part of the higher order completion of the $R^2$-invariant \cite{Baron:2020xel}. What we have shown here is that they cannot come with an independent coefficient. Note that so far our results are consistent with the string $\alpha'$-expansion.

\section{\boldmath \texorpdfstring{$R^4$}{R**4} invariant}\label{sec:R4}
Consider $L=L_4+L_5$ where
\begin{equation}
L_4\sim R^4\,,\qquad L_5\sim\partial^3F^5\,.
\end{equation}
We will show that no $L_5$, which would make $L$ Lorentz invariant up to total derivatives and higher order terms in the fields, exists.

It will be enough to work in the approximation where we neglect all terms involving $F_A$ or a divergence of $F_{ABC}$. We then have the necessary conditions (\ref{eq:cond2-3}). In particular, the first equation says that
\begin{equation}
-G^{\o{ab}\u{de}}F^{\o c}{}_{\u{de}}
-G^{[\o a}\eta^{\o b]\o c}
+3G^{\o{abc}}
+H^{\o{abc}}
\sim\,
\mbox{total derivatives}\,.
\label{eq:cond2-3a}
\end{equation}
Symmetrizing in $\o{bc}$ and neglecting for the moment terms where the index $\o b$ or $\o c$ is sitting on a derivative the last two terms on the LHS drop out and we find the condition
\begin{equation}
-G^{\o a(\o b}{}_{\u{de}}F^{\o c)\u{de}}
-\tfrac12G^{\o a}\eta^{\o b\o c}
+\tfrac12G^{(\o b}\eta^{\o c)\o a}
\sim
\mbox{total derivatives}+\partial^{\o b,\o c}-\mbox{terms}\,,
\label{eq:R4-cond}
\end{equation}
which strongly constrains the possible form of the $R^4$-terms. To see this we note that the $R^4$-terms have eight possible structures
\begin{equation}
L_4=c_1I_1+c_2I_2+\ldots+c_8I_8\,,
\end{equation}
with
\begin{align}
I_1=&\,R_{\o{ab}\u{ef}}R^{\o{ab}\u{ef}}R_{\o{cd}\u{gh}}R^{\o{cd}\u{gh}}\,, & I_5=&\,R_{\o{ab}\u{ef}}R_{\o{cd}}{}^{\u{ef}}R^{\o{bc}}{}_{\u{gh}}R^{\o{da}\u{gh}}\,,\nonumber\\
I_2=&\,R_{\o{ab}\u{ef}}R^{\o{cd}\u{ef}}R^{\o{ab}\u{gh}}R_{\o{cd}\u{gh}}\,, & I_6=&\,R_{\o{ab}\u{ef}}R^{\o{cd}\u{fg}}R^{\o{ab}}{}_{\u{gh}}R_{\o{cd}}{}^{\u{he}}\,,\\
I_3=&\,R_{\o{ab}\u{ef}}R^{\o{ab}\u{fg}}R_{\o{cd}\u{gh}}R^{\o{cd}\u{he}}\,, & I_7=&\,R_{\o{ab}\u{ef}}R^{\o{bc}\u{fg}}R_{\o{cd}\u{gh}}R^{\o{da}\u{he}}\,,\nonumber\\
I_4=&\,R_{\o{ab}\u{ef}}R^{\o{bc}\u{ef}}R_{\o{cd}\u{gh}}R^{\o{da}\u{gh}}\,, & I_8=&\,R_{\o{ab}\u{ef}}R^{\o{bc}\u{fg}}R_{\o{cd}}{}^{\u{he}}R^{\o{da}}{}_{\u{gh}}\,.\nonumber
\end{align}
Taking the derivative of $L_4$ with respect to $R_{\o{ab}\u{cd}}$ gives $G^{\o{ab}\u{cd}}$ and the condition (\ref{eq:R4-cond}) becomes
\begin{align}
&c_4F^{\o c}{}_{\u{de}}\partial^{\o f}F^{\o b\u{de}}\partial^{\o d}F^{\o a\u{gh}}R_{\o{fd}\u{gh}}
+c_4F^{\o c}{}_{\u{de}}\partial^{\o f}F^{\o b\u{gh}}\partial^{\o d}F^{\o a\u{de}}R_{\o{fd}\u{gh}}
+2c_5F^{\o c}{}_{\u{de}}\partial_{\o f}F^{\o b\u{gh}}\partial_{\o d}F^{\o a}{}_{\u{gh}}R^{\o{fd}\u{de}}
\nonumber\\
&{}
+2c_7F^{\o c}{}_{\u{de}}\partial^{\o f}F^{\o b\u{eg}}\partial^{\o d}F^{\o a\u{hd}}R_{\o{fd}\u{gh}}
+c_8F^{\o c}{}_{\u{de}}\partial^{\o f}F^{\o b\u{eg}}\partial^{\o d}F^{\o a}{}_{\u{gh}}R_{\o{fd}}{}^{\u{hd}}
+c_8F^{\o c\u{de}}\partial^{\o f}F^{\o b\u{gh}}\partial^{\o d}F^{\o a}{}_{\u{eg}}R_{\o{fd}\u{hd}}
\nonumber\\
&{}
-\tfrac14G^{\o a}\eta^{\o b\o c}
+\tfrac14G^{\o b}\eta^{\o c\o a}
+(\o b\leftrightarrow\o c)
\sim
\mbox{total derivatives}+\partial^{\o a,\o b,\o c}-\mbox{terms}\,,
\end{align}
where for simplicity we ignored also terms involving $\partial^{\o a}$. This condition requires that the derivative can be taken out of the first two factors (recall that we are dropping divergences of $F$), which in turn requires
\begin{equation}
c_4=2c_5\,,\qquad
c_8=2c_7\,.
\end{equation}
We also find $G^{\o a}\sim0$. Repeating the calculation but this time keeping track of the $\partial^{\o a}$-terms one finds the additional conditions
\begin{equation}
c_5=-4c_1=-2c_2\,,\qquad
c_7=-2c_3=-4c_6\,.
\end{equation}
Clearly we will find similar conditions when we look at the $\lambda^{(-)}$ variation and therefore we should impose that $L_4$ is symmetric under exchanging underlined and overlined indices. This requires $c_6=-4c_1$ and leads to a unique expression
\begin{align}
L_4
=&\,
R_{\o{ab}\u{ef}}R^{\o{bc}\u{fg}}R_{\o{cd}\u{he}}R^{\o{da}\u{gh}}
+\tfrac12R_{\o{ab}\u{ef}}R^{\o{bc}\u{fg}}R_{\o{cd}\u{gh}}R^{\o{da}\u{he}}
-\tfrac14R_{\o{ab}\u{ef}}R^{\o{ab}\u{fg}}R_{\o{cd}\u{gh}}R^{\o{cd}\u{he}}
\nonumber\\
&{}
-\tfrac14R_{\o{ab}\u{ef}}R^{\o{bc}\u{ef}}R_{\o{cd}\u{gh}}R^{\o{da}\u{gh}}
-\tfrac18R_{\o{ab}\u{ef}}R_{\o{cd}}{}^{\u{ef}}R^{\o{bc}}{}_{\u{gh}}R^{\o{da}\u{gh}}
-\tfrac18R_{\o{ab}\u{ef}}R^{\o{cd}\u{fg}}R^{\o{ab}}{}_{\u{gh}}R_{\o{cd}}{}^{\u{he}}
\nonumber\\
&{}
+\tfrac{1}{16}R_{\o{ab}\u{ef}}R^{\o{cd}\u{ef}}R^{\o{ab}\u{gh}}R_{\o{cd}\u{gh}}
+\tfrac{1}{32}R_{\o{ab}\u{ef}}R^{\o{ab}\u{ef}}R_{\o{cd}\u{gh}}R^{\o{cd}\u{gh}}\,,
\end{align}
up to an overall coefficient. Remarkably, ignoring the difference between the over-/underlined indices, this is precisely the form of the $R^4$-terms in the string effective action (\ref{eq:alpha3correction}), namely
\begin{equation}
L_4=\frac{1}{3\cdot2^7}\o t_8\u t_8R^4\,,
\end{equation}
with $\o t_8$ ($\u t_8$) defined as in (\ref{eq:t8}) with over(under)lined indices. Note that the $\varepsilon_8\varepsilon_8$-terms are absent, but this is consistent since these terms are a total derivative at the leading order in fields. At this stage things seem very promising but we are not done yet.

Going back to (\ref{eq:cond2-3a}) we can now extract
\begin{equation}
G^{\o{abc}}\sim t^{\u d_1\cdots\u d_8}F^{[\o a}{}_{\u d_1\u d_2}K^{\o{bc}]}{}_{\u d_3\cdots\u d_8}+\partial^{[\o a}W\partial^{\o b}X\partial^{\o c]}YZ+\mbox{total derivatives}
\end{equation}
for some $W,X,Y,Z$ where
\begin{align}
K^{\o{ab}}{}_{\u c_1\cdots\u c_6}
=&
\tfrac{3}{32}R^{\o{de}}{}_{\u c_1\u c_2}R_{\o{de}\u c_3\u c_4}R^{\o{ab}}{}_{\u c_5\u c_6}
-\tfrac38R^{\o{ad}}{}_{\u c_1\u c_2}R^{\o{be}}{}_{\u c_3\u c_4}R_{\o{de}\u c_5\u c_6}
\nonumber\\
&{}
-\tfrac38\partial^{\o a}F^{\o d}{}_{\u c_1\u c_2}\partial^{\o b}F^{\o e}{}_{\u c_3\u c_4}R_{\o{de}\u c_5\u c_6}
+\tfrac18\partial^{\o d}F^{\o a}{}_{\u c_1\u c_2}\partial^{\o e}F^{\o b}{}_{\u c_3\u c_4}R_{\o{de}\u c_5\u c_6}\,.
\label{eq:K1}
\end{align}
We also find $H^{\o{abc}}$ but all we need to know about these terms is that they do not contain $F_{\o{abc}}$ or $F_{\u{abc}}$, which is clear from the fact that $G^{\o{ab}\u{cd}}$ does not contain these fields.\footnote{There is a small caveat here in that there are ambiguous terms that can cancel between $G^{\o{abc}}$ and $H^{\o{abc}}$. These are the terms involving $W,X,Y,Z$ in $G^{\o{abc}}$. We could then have for example $W=F_{\o{abc}}$ but all we will actually need is that the derivative of $H^{\o{abc}}$ with respect to $F_{\o{efg}}$ is a total derivative, which is indeed the case either way.}

But now we have to worry about satisfying the integrability condition
\begin{equation}
\frac{\partial^2L}{\partial F_{\o d\u{ef}}\partial F_{\o{abc}}}=\frac{\partial^2L}{\partial F_{\o{abc}}\partial F_{\o d\u{ef}}}
\qquad\mbox{or}\qquad
\frac{\partial G^{\o{abc}}}{\partial F_{\o d\u{ef}}}=\frac{\partial G^{\u{ef}\o d}}{\partial F_{\o{abc}}}\,.
\end{equation}
The LHS is $\eta^{\o d[\o a}(\u t_8K)^{\o{bc}]\u{ef}}$ plus total derivatives and, using the second condition in (\ref{eq:cond2-3}) with over(under)lined indices exchanged, the RHS vanishes up to total derivatives. This therefore requires $(\u t_8K)^{\o{ab}\u{ef}}$ to be a total derivative, since otherwise there will be an obstruction to writing the $R^4$ terms in $O(D,D)$ covariant form. As a first check one can verify that the Lorentz variation of $(\u t_8K)^{\o{ab}\u{ef}}$ is indeed a total derivative. It is not very hard to show that
\begin{align}
(\u t_8K)^{\o{ab}\u{ef}}
\sim
-\tfrac32
t^{\u{ef}\u g_1\cdots\u g_6}
\partial^{\o a}F^{\o c}{}_{\u g_1\u g_2}\partial^{\o b}F^{\o d}{}_{\u g_3\u g_4}\partial_{\o c}F_{\o d}{}_{\u g_5\u g_6}
+\mbox{total derivatives}\,.
\end{align}
To see whether the first term is a total derivative we linearize the expression using (\ref{eq:F-lin}) and we get\footnote{Alternatively one can linearize (\ref{eq:K1}) directly. Note that we are dropping terms involving $\partial^2\hat e$, $\partial^{\o a}\hat e_{\o aB}$ or $\partial^{\u a}\hat e_{\u aB}$ which also implies that terms with a contracted derivative $\partial_{\o a}\hat e\partial^{\o a}\hat e\hat e$ vanish up to total derivatives.}
\begin{equation}
(\u t_8K)^{\o{ab}}{}_{\u{ef}}\sim
-48\partial^{[\o a}{}_{[\u e}\hat e_{|\o d\u h|}\partial^{\o b]\o c\u h}{}_{\u f]}\hat e^{\o d\u g}\hat e_{\o c\u g}
+24\partial^{[\o a}{}_{[\u e}\hat e_{|\o d\u h|}\partial^{\o b]\o c\u g}{}_{\u f]}\hat e^{\o d\u h}\hat e_{\o c\u g}
+\ldots\,,
\end{equation}
where the ellipsis stands for terms of higher order and total derivatives. It is clear from the structure that this cannot be written as a total derivative. Therefore there is indeed an obstruction to completing the $O(D,D)$ invariant $R^4$-terms with terms of fifth order in fields. The existence of these terms in the string effective action means that it is not compatible with $O(D,D)$ symmetry.

\section{Conclusions}
We have analyzed possible $O(D,D)$ invariants up to order $\alpha'^3$ (excluding Chern-Simons terms). We have found that the only higher derivative $O(D,D)$ invariant up to this order is of the form Riemann squared. In fact it is known to capture the first $\alpha'$ correction to the bosonic and heterotic string effective actions \cite{Marques:2015vua} and it seems likely that its completion accounts also for the $\alpha'^2$ correction in these theories \cite{Baron:2018lve,Baron:2020xel}. The absence of any $R^4$ invariant at order $\alpha'^3$ is problematic however. Such terms are known to appear in all string theories with the transcendental coefficient $\zeta(3)$. It therefore appears, barring some loophole in our analysis, that the frameworks of DFT/generalized geometry/Exceptional Field Theory\footnote{Note that we have only used basic features of the $O(D,D)$ covariant formulation which are shared by these different duality symmetric formalisms. In particular the same conclusion should follow if one uses the metric-like formulation of DFT rather than the frame-like flux formulation used here. Note also that requiring only $O(d,d)$ with $d<D$ (without restricting the form of the background) does not evade the problems encountered here (as long as say $d>2$).} are not able to account for all the string $\alpha'$ corrections beyond order $\alpha'^2$. Optimistically one could hope that all corrections with non-transcendental coefficients can be cast in an $O(D,D)$ invariant form. Note that this statement applies to the uncompactified theory. Of course, if one compactifies on $T^d$ the $O(d,d)$ symmetry should be there to all orders in $\alpha'$ \cite{Sen:1991zi}. It would be interesting to try to use it to constrain the form of higher $\alpha'$ corrections using similar tools to those used here (see \cite{Eloy:2019hnl,Eloy:2020dko} for a somewhat different approach).

Note that in our analysis we have allowed for a general modification of the double Lorentz transformations. We did not modify the generalized diffeomorphisms. However, this does not lead to any loss of generality since a modification of these necessarily takes the form $\delta' E_A{}^ME_{BM}=Z_{AB}$, for some anti-symmetric $Z_{AB}$, which is equivalent to a modification of the double Lorentz transformations (\ref{eq:deltaE}). The only difference is that the transformation of the generalized dilaton could be non-zero, but this also does not affect our analysis since it leads to terms involving $F_A$, which we were neglecting anyway.

Our results might have important consequences for generalizations of T-duality such as non-abelian and Poisson-Lie T-duality. In the $O(D,D)$ covariant formulation these dualities become very simple (e.g. \cite{Hassler:2017yza}) and this fact has been used recently to find their first $\alpha'$ correction \cite{Borsato:2020wwk,Hassler:2020tvz,Codina:2020yma}. These results extend to any order in $\alpha'$ as long as an $O(D,D)$ covariant formulation exists. Since this does not appear to be the case at order $\alpha'^3$, this argument breaks down at this order. In light of this it would be interesting to investigate the fate of these generalized T-dualities at order $\alpha'^3$. Similar comments apply to closely related integrable deformations of the string sigma model, whose $\alpha'$ correction was found using $O(D,D)$ in \cite{Borsato:2020bqo}. Again our analysis suggests potential problems for these at order $\alpha'^3$. Of course, when we restrict to abelian T-dualities and deformations built using these, there should not be any problem in completing them to all orders in $\alpha'$.

\section*{Acknowledgements}
We thank R. Borsato, D. Marqu\'es and A. Tseytlin for interesting discussions and helpful comments on the manuscript. We also thank M. Krbek for pointing out the reference \cite{Utiyama:1956sy} to us. This work is supported by the grant ``Integrable Deformations'' (GA20-04800S) from the Czech Science Foundation (GA\v CR).

\vspace{2cm}
\appendix

\section{Argument that Lagrangian is expressed in terms of the generalized fluxes}\label{app:L-F}
Here we will argue that generalized diffeomorphism invariance requires the Lagrangian to be expressed in terms of the generalized fluxes. We follow the Utiyama approach \cite{Utiyama:1956sy} used in section \ref{sec:leading}. Under a generalized diffeomorphism, parametrized by $Y_A$, the generalized vielbein transforms as
\begin{equation}
\delta E_A{}^ME_{BM}=2\partial_{[A}Y_{B]}-F_{ABC}Y^C\,.
\end{equation}
When $A$ and $B$ are both over(under)lined this can be absorbed into a double Lorentz transformation, so we may keep only the transformations with say $A=\o a$, $B=\u b$. For simplicity we will ignore the dependence on the generalized dilaton. Defining
\begin{equation}
G_M{}^{A,B_1\cdots B_k}=\frac{\partial L}{\partial(\partial_{B_1\cdots B_k}E_A{}^M)}\,,
\end{equation}
and looking only at the $Y_{\o a}$-terms, the requirement that the Lagrangian be invariant\footnote{We have not allowed for invariance up to total derivatives here, but we expect the end result to be the same also in that case.} is
\begin{align}
&\sum_{k=0}^N\left(
G_M{}^{\o a,B_1\cdots B_k}\partial_{B_1\cdots B_k}(E^{\u bM}[-\partial_{\u b}Y_{\o a}+F_{\u b\o{ac}}Y^{\o c}])
+G_M{}^{\u a,B_1\cdots B_k}\partial_{B_1\cdots B_k}(E^{\o bM}[\partial_{\u a}Y_{\o b}-F_{\u a\o{bc}}Y^{\o c}])
\right)
\nonumber\\
&\qquad=0
\,.
\end{align}
Terms with different numbers of derivatives of $Y_{\o a}$ are independent and looking at the terms with the maximum number of derivatives of $Y_{\o a}$ gives
\begin{equation}
G^{\o a\u b,B_1\cdots B_N}\partial_{B_1\cdots B_N\u b}Y_{\o a}
-G^{\u b\o a,B_1\cdots B_N}\partial_{B_1\cdots B_N\u b}Y_{\o a}
=0\,.
\end{equation}
The part of $G^{\o a\u b,B_1\cdots B_N}$ anti-symmetric in the first two indices can only vanish in the case $N=0$, corresponding to the trivially invariant terms $\eta_{AB}$ and $\mathcal H_{AB}$. For $N>0$ the expression can only vanish because of anti-symmetry in $\u b$ and one of the $B_i$ indices. Looking at the $Y_{\u a}$-terms instead we will find a similar condition with over(under)lined indices exchanged. Together these imply that $G^{AB,B_1\cdots B_N}$ is anti-symmetric in (say) the first three indices. This means that the terms in the Lagrangian with the most derivatives of $E_A{}^M$ are actually of the form $3\partial_{B_2\cdots B_N[B_1}E_A{}^ME_{B]M}=\partial_{B_2\cdots B_N}F_{ABB_1}$. Now we can repeat the same argument for the terms with $N-1$ derivatives of $E_A{}^M$ and so on. Finally, including also the generalized dilaton one finds that it should appear through the generalized flux $F_A$.

\bibliographystyle{nb}
\bibliography{biblio}{}

\end{document}